# A quantum model for the stock market

**Authors:**

Chao Zhang [a,*], Lu Huang [b]

**Affiliations:**

[a] *School of Physics and Engineering, Sun Yat-sen University, Guangzhou 510275, China*

[b] *School of Economics and Business Administration, Chongqing University, Chongqing 400044, China*

**Contact information for the corresponding author:**

E-mail: zhang_chao@126.com

**Abstract:**

Beginning with several basic hypotheses of quantum mechanics, we give a new quantum model in econophysics. In this model, we define wave functions and operators of the stock market to establish the Schrödinger equation for the stock price. Based on this theoretical framework, an example of a driven infinite quantum well is considered, in which we use a cosine distribution to simulate the state of stock price in equilibrium. After adding an external field into the Hamiltonian to analytically calculate the wave function, the distribution and the average value of the rate of return are shown.

## 1. Introduction

The study of econophysics originated in the 1990s [1]. Some physicists found that a few models of statistical physics could be used to describe the complexity of financial markets [2,3]. Nowadays most of the econophysics theories are established on the basis of statistical physics.

Statistical physics is only one branch of physics. After several years' development of econophysics, some physicists began to use other physical theories to study economics. Quantum is one of the most important theories in contemporary physics. It was the first time that quantum theory was applied to the financial markets when someone used quantum field theory to make portfolios as a financial filed [4,5], in which path integral and differential manifold were introduced as the tools to describe the change of financial markets after the gauge transformation. This idea is the same as the essence of the stochastic analysis in finance. There are some other interesting quantum models. For instance, Schaden originally described assets and cash hold by the investor as a wave function to model the financial markets, which was different from usual financial methods using the change of the asset price to be the description [6]. In addition, people paid more attention to the quantum game theory and that was useful in the trading strategies [7,8].

In recent years, an increasing number of quantum models were applied to finance [9-15], which attracted great attention. In this paper, we start from a new approach to explore the quantum application to the stock market. In Section 2 we begin from several basic hypotheses of quantum mechanics to establish a new quantum finance model, which can be used to study the dynamics of the stock price. In Section 3 a simple Hamiltonian of a stock is given. By solving the corresponding partial differential equation, we quantitatively describe the volatility of the stock in Chinese stock market under the new framework of quantum finance theory. A conclusion is illustrated in Section 4.

## 2. The quantum model

Quantum mechanics is the theory describing the micro-world. Now this theory is to be applied in the stock market, in which the stock index is based on the statistics of the share prices of many representative stocks. If we regard the index as a macro-scale object, it is reasonable to take every stock, which constitutes the index, as a micro system. The stock is always traded at certain prices, which presents its corpuscular property. Meanwhile, the stock price fluctuates in the market, which presents the wave property. Due to this wave-particle dualism, we suppose the micro-scale stock as a quantum system. Rules are different between the quantum and classical mechanics. In order to describe the quantum characters of the stock, we are going to build a price model on the basic hypotheses of quantum mechanics.

### *2.1. State vector in the Hilbert space*

In the first hypothesis of quantum mechanics, the vector called "wave function" in the Hilbert space describes the state of the quantum system. Being different from previous quantum finance model [6,9], here we take the square modulus of the wave function $\psi(\wp,t)$ as the price distribution, where $\wp$ denotes the stock price and $t$ is the time. Due to the wave property of the stock, the wave function in the so-called price representation can be expressed by Dirac notations as

$$|\psi\rangle = \sum_n |\phi_n\rangle c_n \ , \tag{1}$$

where $|\phi_n\rangle$ is the possible state of the stock system and the coefficient $c_n = \langle\phi_n|\psi\rangle$. It is exactly the superposition principle of quantum mechanics, which has been studied by Shi [16] and Piotrowski [17] in the stock market.

As a result, the state of the stock price before trading should be a wave packet, or rather a distribution, which is the superposition of its various possible states with different prices. Under the influence of external factors, investors buy or sell stocks at some price. Such a trading process can be viewed as a physical measurement or an observation. As a result, the state of the stock turns to be one of the possible states, which has a certain price, i.e. the trading price. In this case, $|c_n|^2$ denotes the appearance probability of each state. In the statistical interpretation of the wave function, $|\psi(\wp,t)|^2$ is the probability density of the stock price at time $t$, and

$$P(t) = \int_a^b |\psi(\wp,t)|^2 dp \tag{2}$$

is the probability of the stock price between $a$ and $b$ at time $t$.

Actually the fluctuation of the stock price can be viewed as the evolution of the wave function $\psi(\wp,t)$ and we will present the corresponding Schrödinger equation in the following text.

### *2.2. Hermitian operator for the stock market*

In quantum mechanics, physical quantities that are used to describe the system can be written as Hermitian operators in the Hilbert space, which determine the observable states. The values of physical quantities should be the eigenvalues of corresponding operators. While in the stock market, each Hermitian operator represents an economic quantity. For example, the price operator $\hat{\wp}$ (here we approximate the price as a continuous-variable) corresponds to the position operator $\hat{x}$ in quantum mechanics, which has been originally used in the Brownian motion of the stock price [18]. Therefore the fluctuation of the stock price can be viewed as the motion of a particle in the space. Moreover, the energy of the stock, which represents the intensity of the price's movement, can be described by the Hamiltonian that plays a key role in the Schrödinger equation.

### *2.3. Uncertainty principle*

The relation of two variables that do not commute with each other can be demonstrated by the uncertainty principle. For example, the position and the momentum are two familiar conjugate variables in quantum mechanics. The product of their standard deviation is greater than or equal to a certain constant. This means one cannot simultaneously get the accurate values of both position and momentum. The more precisely one variable is measured, the less precisely the other one can be known.

As is mentioned above, the stock price corresponds to the position. Meanwhile there should be another variable $T$ corresponding to the momentum. As guidance in quantum theory, the correspondence principle figures out that when the laws within the framework of the micro-world extend to macro scope, the results should be consistent with the outcomes of the classical laws. In the macro system, the momentum can be written as the mass times the first-order time derivative of the position in some special cases. As a result, in our quantum finance model

$$T = m_0 \frac{d}{dt} \wp \, , \tag{3}$$

where we call the constant $m_0$ “the mass of stock”. $T$ is a variable denoting the rate of price change, which corresponds to the trend of the price in the stock market. In our model, the uncertainty principle thus can be written as

$$\Delta \wp \Delta T \geq \frac{\hbar}{2} , \tag{4}$$

where $\Delta \wp$ and $\Delta T$ are the standard deviations of the price and the trend respectively, and $\hbar$ is the reduced Planck constant in quantum mechanics. The equality is achieved when the wave function of the system is a Gaussian distribution function. Meanwhile, in finance, the Gaussian distribution usually may approximately describe the rate of return of the asset in the balanced market [18]. Taking Yuan (the currency unit in China) as the unit of price in the rest of text, we may estimate the standard deviation of the price as $\Delta \wp = 10^{-3}$ Yuan [19]. When the total variation of the stock price is small, the standard deviation of $\frac{d}{dt}\wp$ in the trend (3) can be approximated as

$$\Delta\left(\frac{d}{dt}\wp\right) = \sqrt{\left\langle \left(\frac{d}{dt}\wp\right)^2 \right\rangle - \left\langle \frac{d}{dt}\wp \right\rangle^2} \approx \sqrt{\left\langle \left(\frac{d}{dt}\wp\right)^2 \right\rangle} \, . \tag{5}$$

Meanwhile, the average rate of stock price change can be evaluated as $10^{-2}$ Yuan per ten seconds in Chinese stock market. Via the uncertainty principle (4) we estimate the magnitude of $m_0$ is about $10^{-28}$. Although the unit of this “mass”, which contains units of mass, length and currency, is different from the real mass, it does not affect the calculation of the wave function which is non-dimensional and we still call it mass in this paper. It should be an intrinsic property and represents the inertia of a stock. When the stock has a bigger mass, its price is more difficult to change. In general, stocks having larger market capitalizations, always move slower than the smaller market capitalizations ones. Thus, the mass of the stock may be considered as a quantity representing the market capitalization.

The uncertainty principle (4) can be often seen in finance. For example, at a certain time someone knows nothing but the exact price of a stock. As a result, he certainly does not know the rate of price change at next time and the direction of the price’s movement. In other words, the uncertainty of the trend seems to be infinite. However in the real stock market, we know more than the stock price itself at any time. We can always get the information about how many buyers and sellers there are near the current price (e.g. investors in China are able to see five or ten bid and ask prices and their volumes on the screen via stock trading software). It is actually a distribution of the price within a certain range instead of an exact price. As a result, we can evaluate a standard deviation of the price. Thus the trend of the stock price may be partly known via the uncertainty principle (4). For example, a trader sees the number of buyers is far more than the number of sellers near the current price, he may predict that the price will rise at next time.

In finance, the standard deviation of the asset price is usually an indicator of the financial risks. Introducing the uncertainty principle of quantum theory may be helpful in the study of the risk management theory.

### *2.4. Schrödinger Equation*

With the assumptions of the wave function and the operator above, let us consider a differential equation to calculate the evolution of the stock price distribution over time. In quantum theory, it should be the Schrödinger equation which describes the evolution of the micro-world. Corresponding to our model, it can be expressed as

$$i\hbar \frac{\partial}{\partial t} \psi(\wp, t) = \hat{H} \psi(\wp, t) , \tag{6}$$

where the Hamiltonian $\hat{H} = \hat{H}(\wp, T, t)$ is the function of price, trend and time. When we know the initial state of the price, by solving the partial differential equation (6), we can get the price distribution at any time in the future. The difficulty here is the construction of the Hamiltonian because there are lots of factors impacting the price and the trend of the stock, such as the economic environment, the marketing information, the psychology of investors, etc. It is not easy to quantify them and import them into an operator. Next we will use the theories above to construct a simple Hamiltonian to simulate the fluctuation of the stock price in Chinese market under an ideal periodic impact of external factors.

### 3. The stock in an infinite high square well

In Chinese stock market there is a price limit rule, i.e. the rate of return in a trading day cannot be more than $\pm 10\%$ comparing with the previous day's closing price, which is applied to most stocks in China. This leads us to simulate the fluctuation of the stock price between the price limits in a one-dimensional infinite square well.

We consider an infinite well with width $d_0 = \wp_0 \times 20\%$, where $\wp_0$ is the previous day's closing price of a stock. After the transformation of coordinate

$$\wp' = \wp - \wp_0, \tag{7}$$

the quantum well becomes a symmetric infinite square well with width $d_0$, and the stock price is transformed into the absolute return. Go on and let

$$r = \frac{\wp'}{\wp_0}, \tag{8}$$

thus the variable of coordinate turns to be the rate of the return. At the same time, the width of the well becomes $d = 20\%$. If we approximately take the average stock price in China to be $10$ Yuan, to evaluate the mass of the stock again in the new coordinate system, there should be a division by $10^2$ Yuan from the left side of the uncertainty principle (4). Therefore the new mass denoted by $m$ should be approximately evaluated as $10^{-30}$, in which the dimension of currency vanishes and this is delighted in physics.

Usually, when the market stays in the state of equilibrium, the return distribution can be described approximately by the Gaussian distribution [18] or more precisely by the Lévy distribution [20]. While in Chinese stock market with the price limit rule, the distribution may be more complicated due to the boundary conditions. However, here we only want to quantitatively describe the volatility of the stock return by the quantum model, so we choose a cosine-square function, whose shape is close to the Gaussian distribution, to approximately simulate the return distribution in equilibrium. The reason for such a selection is the ground state of the symmetric infinite well discussed above is a concave cosine function with no zero except for two extreme points [21], which is usually denoted by

$$\psi_0(r) = \sqrt{\frac{2}{d}} \cos(\frac{\pi r}{d}) \tag{9}$$

with corresponding eigenenergy

$$E_0 = \frac{\hbar^2 \pi^2}{2md^2}. \tag{10}$$

According to the property of the wave function, the square modulus of the state (9) is the probability density of the distribution of the rate of the return. As is shown in Fig. 1 (a), what is the same as the Gaussian distribution or the Lévy distribution is that in the center of the well, which corresponds to zero return, the probability density has the maximal value and it decreases symmetrically and gradually towards the left and right sides. The main reason why this distribution is not precise enough is it does not have the fat tails and the sharp peak. The cosine distribution function equals to zero at $r = \pm\ d/2$ due to the boundary condition of the infinite square well. Judging from the shape, the cosine distribution is a good approximation for the Gaussian distribution with a large variance or the Lévy distribution with a great parameter $\alpha$ .

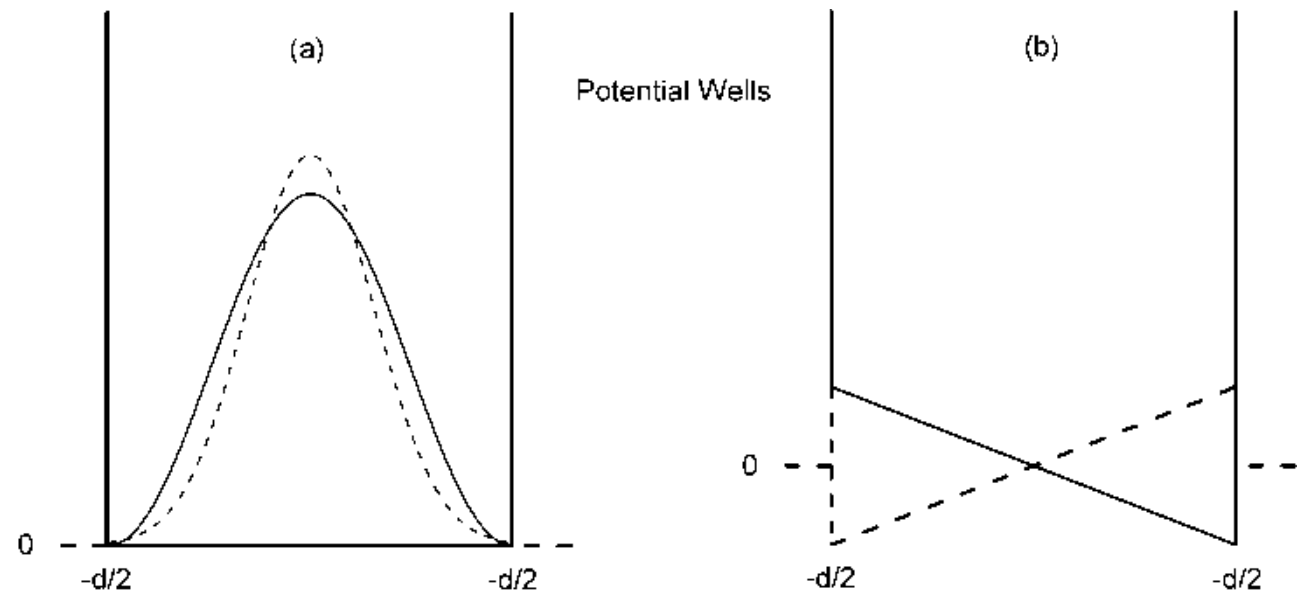

**Fig. 1.** (a) In the infinite square well with width $d = 0.2$, the probability density of the rate of stock return can be approximately described by the Gaussian distribution (dashed line). Here we use a cosine-square function (solid line) to simulate the distribution of the rate of return in equilibrium. (b) When a periodic field put into the Hamiltonian, the bottom of the quantum well begins to slope and it changes periodically.

There is always a lot of market information affecting the stock price. The total effect of information appearing at a certain time is usually conducive either to the stock price's rise or to the stock price's decline. In order to describe the evolution of the rate of return under the information, we consider an idealized model, in which we assume two types of information appear periodically. We may use a cosine function $\cos\omega t$ to simulate the fluctuation of the information, where $\omega$ is the appearance frequency of different kind of information and here we assume $\omega = 10^{-4} s^{-1}$. That means the information fluctuates in a single cycle of about four trading days in China, which may be reasonable. The value of the function $\cos\omega t$ changes between [-1,1] over time, where the information is advantageous to the price's rise when the cosine function is less than zero, and advantageous to the price's decline when it is greater than zero. The stock here is similar to a charged particle moving in the electromagnetic field, where the difference is that the external field of stock market is constructed by the information. The stock price may be influenced by such a field. Under the dipole approximation, the potential energy of the stock can be similarly expressed as $eFr\cos\omega t$, where $e$ is a constant with the same order of magnitude as an elementary charge, and $F$ denotes the amplitude of the external field. The Hamiltonian of this coupled system can be written as

$$\hat{H} = -\frac{\hbar^2}{2m}\frac{\partial^2}{\partial r^2} + eFr\cos\omega t\ , \tag{11}$$

where the first term is the kinetic energy of the stock return, which represents some properties of the stock itself. The second one corresponding to the potential energy reflects the cyclical impact the stock feels in the information field. In order to observe the characters of this system entirely, we make the magnitude difference between the kinetic energy (i.e. ground state energy of the infinite square well without external field) and the potential energy not great, thus estimate $F = 10^{-19}$ here. Because the second term of Hamiltonian contains the rate of return $r$, tilt appears at the bottom of the infinite square well, which is shown in Fig. 1 (b). The slope of the bottom fluctuates periodically due to the change of $\cos\omega t$ over time, which makes the well no longer symmetric. This reflects the imbalance of the market, and the distribution of the rate of return becomes symmetry breaking from the state (9) in equilibrium.

After constructing such an ideal Hamiltonian (11), we need to get the solution of the corresponding Schrödinger equation

$$i\hbar\frac{\partial}{\partial t}\psi(r,t) = \left[-\frac{\hbar^2}{2m}\frac{\partial^2}{\partial r^2} + eFr\cos\omega t\right]\psi(r,t). \tag{12}$$

However, the analytical solution for the similar equation has been studied before [22]. Therefore we just have a brief review. At first, let

$$\psi(r,t) = \varphi(\xi,t)\chi(r,t) \tag{13}$$

with variable substitution

$$\xi = r - \frac{eF\cos\omega t}{m\omega^2}. \tag{14}$$

In the wave function (13), let

$$\chi(r,t) = \exp\left[-\frac{iE_c t}{\hbar} - \frac{ieFr\sin\omega t}{\hbar\omega} - \frac{ie^2F^2(2\omega t - \sin 2\omega t)}{8\hbar m\omega^3}\right], \tag{15}$$

where $E_c$ denotes the energy of the driven system. After both sides of Eqs. (12) are divided by $\chi(r,t)$, we find $\varphi(\xi,t)$ satisfies the time-dependent Schrödinger equation

$$-\frac{\hbar^2}{2m}\frac{\partial^2}{\partial\xi^2}\varphi(\xi,t) - E\varphi(\xi,t) = i\hbar\frac{\partial}{\partial t}\varphi(\xi,t) \tag{16}$$

and its solution can be written as

$$\varphi(\xi,t) = \sum_{l=-\infty}^{\infty} A_l \exp\left(\pm\frac{i\sqrt{2mE}\xi}{\hbar}\right)\exp(-il\omega t), \tag{17}$$

where the energy may be expressed as

$$E = E_c \pm l\hbar\omega. \tag{18}$$

The centric energy $E_c$ is determined by the boundary conditions and close to the energy (10) of the ground state of the stationary Schrödinger equation without external impact. The energy splits around $E_c$ with the integral multiple of energy unit $\hbar\omega$ and every new energy corresponds to a possible state of the system. From Eqs. (13), (15) and (17), the wave function of the rate of stock return can be written as the superposition of all those states

$$\psi(r,t)=\exp\left[-\frac{iE_c t}{\hbar}-\frac{ieFr\sin\omega t}{\hbar\omega}-\frac{ie^2F^2(2\omega t-\sin 2\omega t)}{8\hbar m\omega^3}\right]\times$$
$$\sum_{l=-\infty}^{\infty}A_l\exp(-il\omega t)\left\{\exp\left[ik_l(r-\frac{eF\cos\omega t}{m\omega^2})\right]+(-)^l\exp\left[-ik_l(r-\frac{eF\cos\omega t}{m\omega^2})\right]\right\}, \quad (19)$$

where the wave vector is

$$k_l=\sqrt{2m(E_c+l\hbar\omega)/\hbar^2}\ . \quad (20)$$

In the wave function (19), $A_l$ denotes the amplitude for each possible state, which are some constants to be determined. In Eqs. (19), we use Fourier expansion for the exponential function

$$\exp\left(i\frac{k_l eF\cos\omega t}{m\omega^2}\right)=\sum_{n=-\infty}^{\infty}i^n J_n\left(\frac{k_l eF}{m\omega^2}\right)\exp(in\omega t), \quad (21)$$

where $J_n\left(\frac{k_l eF}{m\omega^2}\right)$ is the $n$ th Bessel function and let Eqs. (19) satisfy the boundary condition of the infinite quantum well

$$\psi(\frac{d}{2},t)=\psi(-\frac{d}{2},t)=0\,. \quad (22)$$

According to physics, this must be satisfied at any time. By the orthogonality of Bessel function [23]

$$\sum_{l=-\infty}^{\infty}J_{n-l}(u)J_{m-l}(u)=\delta_{m,n}\,, \quad (23)$$

the summation over $n$ in Eqs. (21) can be reduced, and Eqs. (19) at the boundary turns to be

$$\sum_{l=-\infty}^{\infty}(-i)^l A_l\left[\exp(ik_l d/2)+(-)^l\exp(-ik_l d/2)\right]J_{n+l}\left(\frac{k_l eF}{m\omega^2}\right)=0\,. \quad (24)$$

By defining two parameters

$$v=\frac{\hbar\omega}{E_c}$$
$$q=\frac{k_0 eF}{m\omega^2} \quad (25)$$

and expanding the wave vector $k_l=k_0\sqrt{1+lv}$ to second order of $v$, the non-normalized approximate solution of $A_l$ may be found [22]

$$A_l = i^l \begin{Bmatrix} J_l(q) + \frac{q(q^2-\pi^2)v^2}{64}\left[J_{l+1}(q) - J_{l-1}(q)\right] - \frac{3q^2 v}{32}\left[J_{l+2}(q) - J_{l-2}(q)\right] \\ -\frac{q^2v^2}{32}\left[J_{l+2}(q) + J_{l-2}(q)\right] - \frac{q^3v^2}{64}\left[J_{l+3}(q) - J_{l-3}(q)\right] \\ +\frac{9q^4v^2}{2048}\left[J_{l+4}(q) + J_{l-4}(q)\right] \end{Bmatrix}, \tag{26}$$

which meets

$$k_0 d = \frac{\pi}{\sqrt{1+q^2v^2/8}}. \tag{27}$$

From Eqs. (20), (25) and (27), $A_l$ can be finally determined. According to Eqs. (19) and (26) we have a numerical calculation for the wave functions at different points in time and simulate their distributions in Fig. 2. When $t=0$, $1000$s and $25000$s, the distributions of the rate of return are plotted with the solid line, the dotted line and the dashed line respectively. At the beginning of $t=0$, the distribution of the rate of return is nearly symmetric with previous day's closing price, which corresponds to the initial state of the stock. The zero return point $r=0$ is the most probable and the probability density decreases towards both sides of the price limits. The average rate of return equals zero at this time. By adding the external field of information in, the distribution of the rate of return starts its evolution over time. When $t=1000$s and $t=25000$s, the maxima of the probability density shift from the point $r=0$, but the peak values seem not to change.

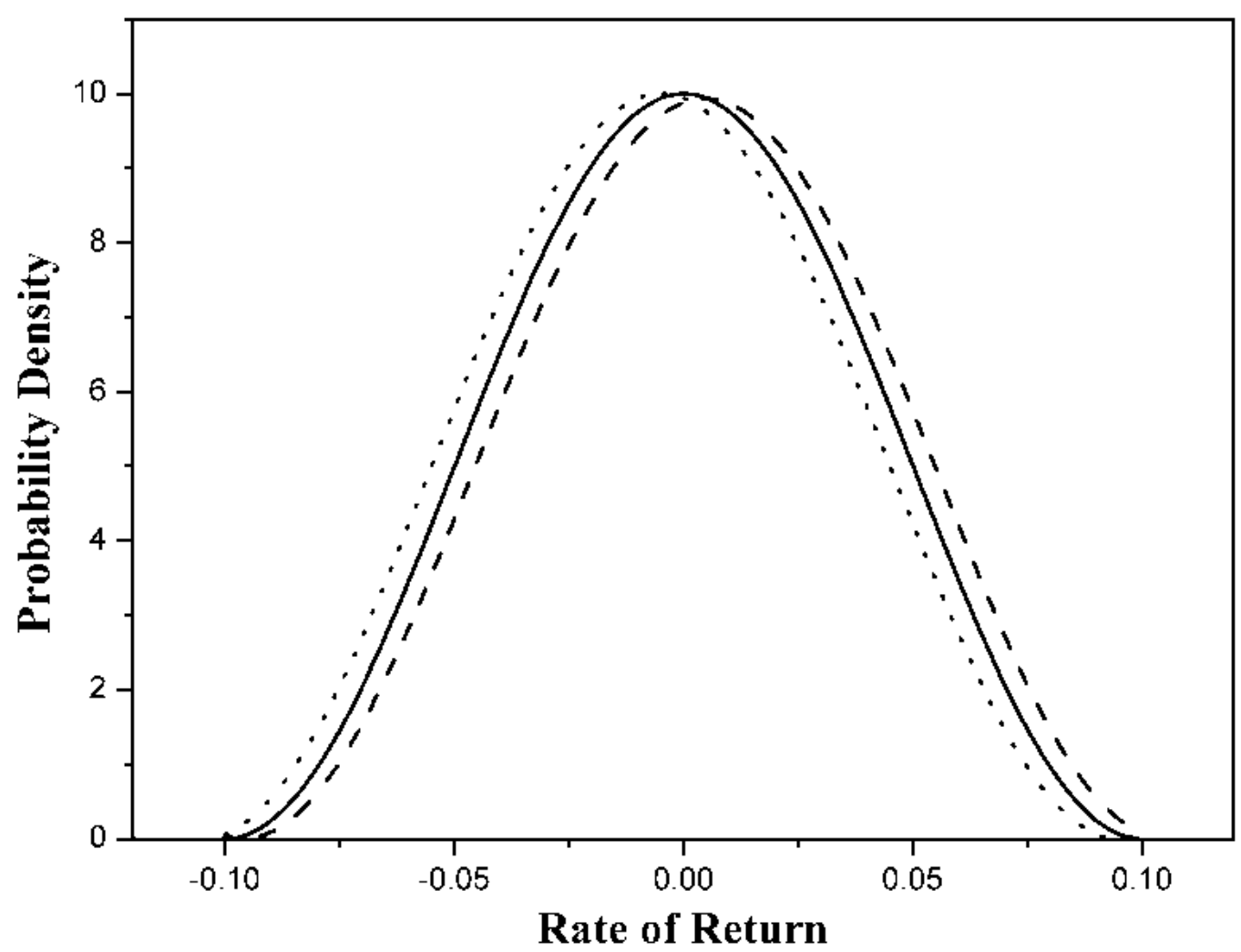

**Fig. 2.** Numerical simulations of the probability density of the rate of return at (a) $t=0$ (solid line), (b) $t=1000$s (dotted line) and (c) $t=25000$s (dashed line), in which parameters are respectively $e=10^{-19}$, $m=10^{-30}$, $\omega=10^{-4}$, $F=10^{-19}$ and $d=0.2$. At $t=0$ the distribution corresponds to the initial state of the system. The external field makes the distributions imbalance at $t=1000$s and $t=25000$s.

The change of the distribution reflects the imbalance of the market under the influence of external information. It also can be seen from the evolution of the average rate of return over time. In quantum mechanics, the average value of the rate of return can be written as

$$\langle r(t)\rangle = \frac{1}{C}\int_{-d/2}^{d/2}\psi^*(r,t)r\psi(r,t)dr, \tag{28}$$

where $C$ is a constant to keep the wave function normalized. The average rate of return fluctuating in one period is shown in Fig. 3. By analyzing Eqs. (19), we obtain that the period of the wave function is $\tau = 2\pi/\omega$. In the first period, the average rate of return is symmetric with $t=\tau/2$, which is due to the bilateral symmetry of the Hamiltonian (11) or the wave function (19) in one time period. For the parameters selected, the average rate of return vibrates about 20 times within a period, and the amplitude of the fluctuation achieves about $\pm 3\%$. It tells that the stock price dose not definitely rise under the advantageous information, while with the disadvantageous information, the stock price may not surely decline. The price is always volatile and Fig. 3 reflects that the price fluctuates more frequently than the total effect of the market information.

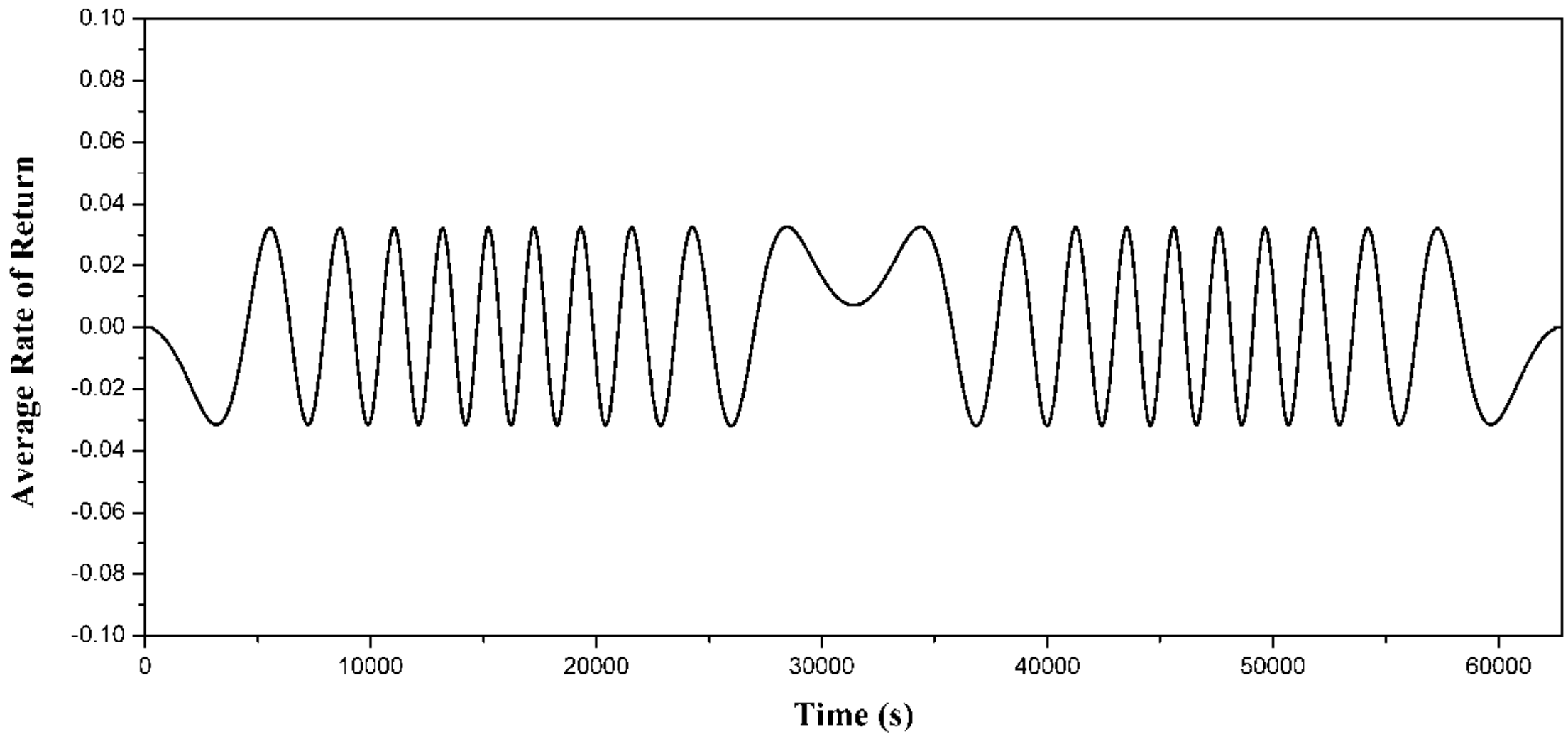

**Fig. 3.** The average rate of return fluctuates in a cycle under the market information with the same parameters as in Fig. 2. The fluctuation of a single cycle is given in the figure, in which the symmetry axis appears at $t=\pi/\omega$ and the amplitude of the fluctuation achieves $\pm 3\%$.

## 4. Conclusions

In this paper, we start from several fundamental assumptions of quantum theory to build a quantum finance model for stock market. By constructing a simple Hamiltonian, we give an example of this quantum model, in which the square modulus of the ground state in the infinite square well represents the distribution of the stock return in equilibrium. By putting an external field in, which simulates the external factors affecting the stock price, we quantitatively describe and discuss the distribution of the rate of return and the evolution of the average rate of return over time. The model gives a new theory of quantum finance, which may be helpful for the development of econophysics. If we can further find some ways to quantify the factors affecting the stock price and write them into the Hamiltonian, or use some more accurate quantum states, such as the ground state of the harmonic oscillator, which exactly equals to the Gaussian distribution, to describe the price distributions in the equilibrium market, the fluctuation of the stock price is able to be estimated more precisely.

## Acknowledgements

We would like to thank Prof. Wang Xuehua for his encouragement of this work and Dr. Liu Jiaming for helpful discussions.